\documentstyle[12pt]{article}
\newcommand{\z}{$Z$}

\newcommand{\swtwo}{$ \sin^{2} \theta_{W} \:$}

\newcommand{\zp}{$Z'$}

\newcommand{\oalf}{\cal{O}($\alpha$)$\:$}

\newcommand{\nobody}{\rule{0ex}{1ex}}
\newcommand{\nn}{\nobody \hfill \\ \noindent }
\begin{document}
%--------------------------------------------------------------
%\input{zetitle}      % Title page
%--------------------------------------------------------------
%----------------------- status 9 nov 91 --------------------
%\begin{titlepage}
\thispagestyle{empty}
\title{
\vspace{-1.4cm}
\begin{flushleft}{\normalsize
{LMU-91/06}
\\
{\tt hep-ph/9808374}
%\\ reconstructed from draft
%\\ status: TR 13-07-98
}\end{flushleft}
\vspace{.5cm} {\LARGE \bf
$ZZ'$ Mixing in Presence of \\ Standard Weak Loop Corrections
%TR\\
%TRand\\ {\tt ZEFIT} - a Package to be Used with {\tt ZFITTER}
\vspace{0.4cm}\\
}}
\author{
{\bf A. Leike}$^{1,2,}$\thanks{Supported in part by
the German Bundesministerium f\"ur Wissenschaft und Technologie}$\;^,$\thanks{
Supported in part by
 the Alexander von Humboldt-Stiftung}$\;$,$\;$
{\bf S. Riemann}$\nobody^{1,*,}$\thanks{e-mail: RIEMANNS@CERNVM}$\;$,$\;$
{\bf T. Riemann}$^{1,3}$
\vspace{1.cm}\\
$^1${\em Institut f\"ur Hochenergiephysik, O-1615 Zeuthen, Germany}
\vspace{0.4cm}\\
$^2${\em Ludwig-Maximilians-Universit\"at, Sektion Physik,}
\\ {\em W-8000 M\"unchen, Germany}
\vspace{0.4cm}\\
$^3${\em Theory Division, CERN, CH-1211 Geneve, Switzerland}
\vspace{0.4cm}\\}
%\date{\today}
\date{December 1, 1991}

\maketitle
%*********************************************
\begin{abstract}
\noindent
We derive a method for a common treatment of $Z'$ exchange,
QED corrections, and weak loops.
It is based on the form factor approach to the description of weak
loop corrections
to partial $Z$ widths and cross sections.
 Problems connected with \z,\zp $\;$ mixing
are discussed with special care.
Our theoretical results are applied to the package {\tt ZFITTER}. 
We
demonstrate two different ways to the data analysis - one based on an
extension of the standard model cross sections, the other on model-independent
formulae together with the $Z$ width calculations in presence of a
$Z^{\prime}$.
\\ With the resulting package {\tt ZEFIT}$\oplus${\tt ZFITTER},
LEP1 data from fermion pair production, including Bhabha scattering,
can be analysed {\it on}, but also {\it off} the $Z$ peak.
Further, the code may be used at very high energies,
e.g. in the region of a possible future linear $e^+e^-$-collider.
\end{abstract}
\vfill\eject
%\end{titlepage}
%------------------------------------------------------------- 
%TR: input done 10/07/1998
%\input{zeintro}      % Introduction    sect.1
%
%------------------------10 nov 1991 -----------------------------------
%
\section{Introduction}
Although the Standard Model \cite{no:gws} has been verified with a precision
including one loop corrections, there is a general consensus that we are
far away from a final understanding of the elementary particle world.
A unification of forces seems to happen at much higher mass scales
than are directly accessible by present accelerators. 
Candidates for a
truly unifying theory usually predict additional gauge degrees of freedom,
thus leading in a natural way to the existence of new, heavy neutral
gauge bosons besides the photon and the $Z$ boson of the standard theory
(see e.g. \cite{673}).
 
\nn
Since $\gamma, Z, Z'$ are neutral particles with vector and axial-vector
couplings,
a search for a $Z'$ is complicated by the fact that there is no
special final state signature. 
Below the production threshold,
theoretical predictions consist of minor quantitative modifications
of the neutral current cross sections.
As a consequence, one needs very precise predictions for cross sections and
asymmetries. 
Important reactions for a
dedicated search are:
\begin{equation}
e^+e^-  \longrightarrow (\gamma, Z, Z') \longrightarrow f^+f^-(\gamma),
\label{ee}
\end{equation}
\begin{equation}
e^+e^-  \longrightarrow (\gamma, Z, Z') \longrightarrow e^+e^-(\gamma),
\label{eeee}
\end{equation}
\begin{equation}
ep \longrightarrow (\gamma, Z, Z') \longrightarrow eX(\gamma).
\label{ep}
\end{equation}
In earlier studies, we investigated QED (i.e. pure photonic)
corrections together with
$Z'$ exchange both in $e^+e^-$-annihilation into fermions \cite{qedzp} and in
$ep$-scattering \cite{Rep}. In the latter reaction, we applied also part
of the material presented here ($Z'$ effects in presence of weak loops without
\z,\zp\ mixing).
 
\nn
In principle, $Z'$ effects can be
searched for by three different effects:  
\\
$\bullet$ via virtual $Z'$ exchange (at sufficient energy, also present
without \z,\zp\ mixing) \\
$\bullet$  via modification of the mass of the standard $Z$ boson seen at LEP1
due to $Z, Z'$ mixing \\
$\bullet$ via modification of the couplings of the standard $Z$ boson seen at
LEP1
due to $Z, Z'$ mixing; this in fact concerns two different, although
related
observables - the \z\ width {\mbox{[$\sim$ peak height]}} and cross
sections {\mbox{[$\sim$ line shape]}}.
 
\nn
For large enough \zp\ masses, the direct cross section contributions due
to
\zp\ exchange may be neglected at LEP1 energies completely. 
Then, one can
concentrate on the consequences of the \z,\zp\ mixing.
{\em In fact, LEP1 is the ideal place to search for this phenomenon.}
 
\nn
{From} existing measurements it is known that the mixing is very small if not
vanishing; see e.g. \cite{zplep} and references quoted therein.
In such a situation, one has to study carefully the interplay of
the weak standard theory loop effects with the \zp $\;$ influence
via particle mixing.
 
\nn
Often the LEP1 data are analysed {\it after} a model-independent interpretation
of the line shape (and of asymmetries) in terms of e.g. partial widths
(or effective couplings),
confronting those as 'observed quantities' with theoretical
expectations.
Our approach will also allow to do so. \\
In addition, we derived formulae
which open the possibility of a direct interpretation of e.g. the line shape
as function of the energy in terms of standard model parameters and, in
parallel, of the \z, \zp\ mixing. (Not mentioning for a moment the direct
\zp\ cross section part.) 
This is done in a specific
way following the form factor notation of weak loop effects on which the
Dubna/Zeuthen approach (and e.g. the code {\tt ZFITTER} which has been 
supplemented here) so heavily relies.
 
\nn
We remind the reader that for each scattering process, there are
four complex-valued form factors $\rho, \kappa_e, \kappa_f, \kappa_{ef}$,
which allow an exact description of the weak (non-photonic) contributions to
the scattering process \cite{15,Rbha}. 
For an exact
description of  \z,\zp\ mixing, one has to understand how these form
factors become modified.
\\ 
Since the partial and total widths of the \z\ boson \cite{zwidth}
contain similar form factors $\rho_Z, \kappa_Z$,
%$\rho_Z, \kappa_Z$ [related
%to some extent, e.g. $\rho$(width) $\sim$ $\rho$(scattering; $s=M_Z^2$)],
a similar procedure has to be applied there, too.
 
\nn
We will also give a short comment on the relation of the formulae derived here
to those used by other authors.
\\ 
This concerns the change of form factors by the \z,\zp\ mixing.
\\ 
Further, we define the weak mixing angle \swtwo\ from the unmixed mass
parameters also in presence of \z,\zp\ mixing.
As a result, the calculation of the standard weak loop corrections remains
unchanged, too. 
Many other authors prefer to redefine the weak mixing angle
as being derived from the physical mass of the \z\ mass eigenstate
\cite{Rdegsir}. 
This leads to the introduction of a $\rho$ parameter in the
definition of the weak mixing angle. 
The difference between the two procedures
is essentially a different book keeping and does not influence the
experimental determination of e.g. the t-quark mass or the \z,\zp\
mixing angle $\theta_{M}$. 
Although, a direct comparison of the
form factors of the different approaches may be meaningless.
 
\nn %---------------------------------------------------------
The present article contains a precise formulation of the
common treatment of standard weak loops and of $Z'$ effects
together with their realisation in the code 
{\tt ZEFIT}${\oplus}${\tt{ZFITTER}}.
Since QED corrections are model-independent in the sense that they are
well-defined if vector and axial vector couplings, mass and width
of the $Z'$ are known, the flexible, semi-analytic multi-purpose code
{\tt ZFITTER} \cite{zfitter} can be used after some modifications  
for the calculation
of QED corrections. 
Thus, we do not discuss explicit formulae on QED
corrections in presence of a $Z'$.
\\ 
In section 3, we comment on the \z\ width.
It is shown how
the weak form factors which re\-nor\-ma\-lise the Fermi constant
($\rho_Z$) and the weak mixing angle ($\kappa_Z$)
are influenced by a mixing of $Z,Z'$.
In section 4, the same is done for the corresponding form factors
of the differential cross sections
$\rho(s,Q^2)$ and $\kappa_e(s,Q^2), \linebreak[0] \kappa_f(s,Q^2),
\linebreak[0] \kappa_{ef}(s,Q^2)$.
Section 5 contains some explicit cross section formulae.
In section 6, the structure of the package {\tt ZEFIT} and its interplay with
{\tt ZFITTER}
is described. 
It contains
the above mentioned changes of form factors together with Born cross
section parts containing \zp\ exchange [for applications at energies
beyond the LEP1 region]. 
Further, in appendix A the explicit vector and axial
vector couplings of the
\zp\ in two important classes of extended gauge models, the
${\rm E}_6$ based and left-right
symmetric models, are determined as functions of one free parameter,
$\theta_{E6}$ or $\alpha_{LR}$, respectively.
Appendix B contains the output of a test program for the use of {\tt ZEFIT}
at LEP1 energies.
\\ 
At the end of this introduction to the subject we would like to stress that
the code originally
is intended for LEP1 physics, but may also be used at higher
energy, e.g. for $Z'$ searches at LEP 200 or LINAC 500.
An application of this kind and a comprehensive list of references may
be found in \cite{zlinac}.
%%TR: \input done 10-7-98
%---------------------------------------------
%The present mass limit for a $Z'$ is $M(Z_2) > 300 - 700$ GeV
%\cite{zelim} depending on the model.
%--------------------------------------------------------------
%\input{zemixing}      % Gauge boson mixing   sect. 2
\section{Gauge Boson Mixing}
%-----------------------------------------------------------------------------
The coupling constants are defined for the symmetry eigenstates \z,
\zp. The Lagrangian
\begin{equation}
{\cal L} = eA_\mu J_\gamma^\mu + g Z_\mu J_Z^\mu + g' Z'_\mu J_{Z'}^\mu
\label{eq31}
\end{equation}
contains currents of the form
\begin{equation}
J_n^\mu = \sum_{f}\;\bar{f} \gamma^\mu\;[v_f(n) + \gamma_5 a_f(n)]\; f, \ \
n = \gamma, Z, Z'.
\label{eq32}
\end{equation}
In the following, complications will arise from a mixing of \z,\zp $\;$
since our renormalisation is performed on mass shell, i.e. for mass eigenstates
$Z_1, Z_2$. In this chapter, we distinguish between these states:
%----------------------------------------------------------------------
\begin{equation}
\left( \begin{array}{c} Z_1 \\ Z_2 \end{array} \right)
=
\left( \begin{array}{rl}  \cos\theta_M & \sin\theta_M \\
                        - \sin\theta_M & \cos\theta_M \end{array} \right)
\left( \begin{array}{c} Z \\ Z' \end{array} \right).
\label{zzmix}
\end{equation}
%----------------------------------------------------------------------
The weak mixing angle
%TR and the $Z, Z'$ mixing angle are
$\theta_W$
is
related to the gauge boson masses 
and to the gauge boson mixing angle $\theta_M$ 
as follows:
\begin{eqnarray}
 t_M^2 = \tan^2 \theta_M = \frac{M_Z^2-M_1^2}{M_2^2-M_Z^2},
\end{eqnarray}
\begin{eqnarray}
M_Z \equiv \frac{M_W}{\cos\theta_W}.
\end{eqnarray}
These equations correspond to \cite{Rdegsir}:
\begin{eqnarray}
\rho_{mix} &=& \frac{M_W^2}{M_1^2\cos^2\theta_W} = \frac{M_Z^2}{M_1^2}
= \frac{1+t_M^2 \, M_2^2/M_1^2}{1+t_M^2} 
= 1 + \sin^2 \theta_M \left( \frac{M_2^2}{M_1^2} -1 \right).
\label{rhomix}
\end{eqnarray}

The Z boson mass measured at LEP1 is $M_1=91.177$ GeV. The couplings
of the mass eigenstates to fermions are :
\begin{equation}
v_f(1) = \cos \theta_M v_f + \frac{g'}{g} \sin \theta_M v'_f, \\
\end{equation}
\begin{equation}
v_f(2) = \cos \theta_M v'_f - \frac{g}{g'} \sin \theta_M v_f, \\
\end{equation}
\begin{equation}
g = (\sqrt{2} G_{\mu} M_1^2)^{1/2}, \hspace{1cm}
v_f = a_f (1-4 |Q_f| \sin^2 \theta_W),\hspace{1cm}
a_f = I_3^L(f),
\end{equation}
with analogue definitions for the axial couplings. \\
The photon couplings are defined such that $Q_e=-1$.
%TR:  10-7-98----------------------------------------------------------
%\input{zewidth}      % Z partial widths    sect. 3
\section{Partial $Z$ widths in Presence of $Z,Z'$ Mixing}
%--------------------------------------------------------------------------
Without $Z,Z'$ mixing, the matrix element
for the decay of the $Z$ boson into a fermion pair
may be written as follows \cite{zwidth}:
%-----
\begin{eqnarray}
{\bar{\cal M}}_f &\sim& 
\sqrt{\frac{G_{\mu}}{\sqrt{2}} M_Z^2}
%G_{\mu}^{\frac{1}{2}} 
\epsilon^{\alpha} \sqrt{\rho_f}
  a_f
 {\bar u} \left[ \gamma_{\alpha} (1+\gamma_5) - 4 \sin^2\theta_W \kappa_f
 \right] u
\nonumber \\
 &\sim& 
\sqrt{\frac{G_{\mu}}{\sqrt{2}} M_Z^2}
%G_{\mu}^{\frac{1}{2}} 
\epsilon^{\alpha} {\bar a}_f
 {\bar u} \left[ \gamma_{\alpha} \gamma_5 + \gamma_{\alpha}
\frac { \bar{v}_f} { \bar{a}_f}
 \right] u,
\label{zmate}
\end{eqnarray}
%-----
where $\rho$ and $\kappa$ contain the weak loop corrections as determined
in the on mass shell renormalisation scheme,
%new here:
after the replacement of the coupling constant $\alpha$ by the muon
decay constant. 
They are related by:
%-----
\begin{eqnarray}
\frac{\pi\alpha}{2\sin^2\theta_W\cos^2\theta_W}
&=&
\frac{G_{\mu}}{\sqrt{2}}M_Z^2(1-\Delta r).
\end{eqnarray}
%-----
The factor $(1-\Delta r)$ becomes part of the form factor $\rho$
\cite{zwidth}. 
\\
The resulting decay width is:
%-----
\begin{eqnarray}
{\bar \Gamma}_f &=& 
\frac{G_{\mu}}{\sqrt{2}} \frac{M_Z^3}{12 \pi} 
c_f  \rho_f
           \left[ 1 - 4 |Q_f| \sin^2\theta_W \kappa_f
                    + 8 (|Q_f| \sin^2\theta_W \kappa_f)^2 \right]
\nonumber \\
         &=& \frac{G_{\mu}}{\sqrt{2}} \frac{M_Z^3}{6 \pi} c_f
             \left[ {\bar v}^2 + {\bar a}^2  \right],
\label{zwidth}
\end{eqnarray}
%-----
where $c_f$ is a color factor in case of quarks.
Further, we used that
%-----
\begin{equation}
\frac{\bar{v}_f}{\bar{a}_f} = 1 - 4|Q_f|\sin ^2 \theta_W \kappa_f.
\label{ratva}
\end{equation}
%-----
The effective vector and axial vector couplings are defined as follows:
\begin{eqnarray}
\bar a_f = \sqrt{\rho_f} \: I_3^L (f),
\label{axial}
\end{eqnarray}
\begin{eqnarray}
\bar v_f =\bar a_f\left[ 1 - 4 |Q_f |\sin^2 \theta_W
                                        \kappa_f
                    \right],
\end{eqnarray}
where $I_3^L (f)$ is the weak isospin of fermion $f$.
It has to be stressed here that the weak form factors are introduced
originally for the overall
normalisation of the matrix element and for the weak mixing angle, but not for
the axial and vector couplings.
 
\nn
If now $Z$ and $Z'$ mix, the above formulae must be written down for the
mass eigenstate $Z_1$ with mass $M_1=91.177$ GeV and in terms of the couplings
of the 'physical $Z$ boson' $Z_1$. One can rewrite the Born
couplings after mixing as follows:
%-------------------------
\begin{eqnarray}
a_f(1) &=& c_M \; a_f + s_M \frac{g'}{g} \; a_f' 
= (c_M + s_M \frac{g'a'_f}{ga_f} ) a_f,   
\\
a_f(1) &=&   (1-y_f) a_f,
\end{eqnarray}
%-------------------------
\begin{eqnarray}
y_f = - s_M \frac{g'a_f'}{ga_f}  + (1 - c_M) 
\sim - s_M \frac{g'a_f'}{ga_f}.
\end{eqnarray}
%-------------------------
Similarly, we make the ansatz:
%----
\begin{equation}
\frac{v_f(1)}{a_f(1)} = \frac{v_f + t_M v'_f g'/g}
                             {a_f + t_M a'_f g'/g} \equiv
1 - 4|Q_f|\sin ^2 \theta_W  (1- x_f),
\end{equation}
%----
from which we derive:
%
%next two equations were wrong. AL, corr. by TR 19-08-1998:
%
%----
%\begin{equation}
%x_f =     (1-v_f/a_f)^{-1}
%\left( \frac{v_f+t_Mv'_fg'/g} {a_f+t_Ma'_fg'/g} \right),
%\end{equation}
%----
%\begin{equation}
%x_f = \frac{t_f}{1+t_f} \frac{ v'_f/a'_f - v_f/a_f }
%                              { 1 - v_f/a_f     }
%\sim s_M \frac{g'}{g} \frac{a'_f}{a_f} \frac{v'_f/a'_f-v_f/a_f}{v_f/a_f},
%\hspace{1cm}
%t_f = t_M \frac{g'a'_f}{g a_f} \sim s_M \frac{g'a'_f}{g a_f}.
%\end{equation}
%----
%----
\begin{equation}
x_f =     (1-v_f/a_f)^{-1}
\left( \frac{v_f+t_Mv'_fg'/g} {a_f+t_Ma'_fg'/g} 
-\frac{v_f}{a_f}
\right),
\label{corrform}
\end{equation}
\begin{equation}
x_f 
\approx 
s_M\frac{g'}{g}\frac{a'_f}{a_f}\frac{v'_f/a'_f-v_f/a_f}{1-v_f/a_f}.
\end{equation}
%----
In terms of these variables, the Born matrix element after mixing is:
%----
\begin{eqnarray}
{\cal M}_f(1)
 &\sim& 
%G_{\mu}^{\frac{1}{2}} 
 \sqrt{\frac{G_{\mu}}{\sqrt{2}}M_1^2}
\epsilon^{\alpha} {\bar a}_f (1)
 {\bar u} \left[ \gamma_{\alpha} \gamma_5 + \gamma_{\alpha}
\frac { \bar{v}_f(1)} { \bar{a}_f(1)}
 \right] u
\nonumber \\
 &\sim& 
%G_{\mu}^{\frac{1}{2}} 
 \sqrt{\frac{G_{\mu}}{\sqrt{2}}M_1^2}
\epsilon^{\alpha} (1- y_f)
  a_f
 {\bar u} \left[ \gamma_{\alpha} (1+\gamma_5) - 4 \sin^2\theta_W
(1-x_f)
\gamma_{\alpha}  \right] u.
\label{zpmate}
\end{eqnarray}
%-----
Starting from this expression, it is evident how to take into account the
weak loop corrections of the (unmixed) standard theory. The following
replacements have to be performed:
\begin{equation}
\rho_f \rightarrow \rho_f^M = \rho_{mix} (1-y_f)^2 \rho_f,
\end{equation}
\begin{equation}
\kappa_f \rightarrow \kappa_{f}^M = (1-x_f) \kappa_{f},
\end{equation}
%----------------------
%
\begin{eqnarray}
\bar a_f(1) = \sqrt{\rho^M_f} \: I_3^L (f),
\label{axial3}
\end{eqnarray}
\begin{eqnarray}
\bar v_f(1) =\bar a_f\left[ 1 - 4 |Q_f |\sin^2 \theta_W
                                        \kappa^M_f  \right].
\end{eqnarray}

The width of the 'physical $Z$' in presence of standard weak loops and
mixing now is:
%-----
\begin{eqnarray}
{\bar \Gamma}(1)_f &=& \frac{G_{\mu}}{\sqrt{2}} \frac{M_1^3}{12 \pi} c_f
\rho_f^M
           \left[ 1 - 4 |Q_f| \sin^2\theta_W \kappa_f^M
                    + 8 (|Q_f| \sin^2\theta_W \kappa_f^M)^2 \right]
\nonumber \\
         &=& \frac{G_{\mu}}{\sqrt{2}} \frac{M_1^3}{6 \pi} c_f
             \left[ {\bar v}_f(1)^2 + {\bar a}_f(1)^2  \right].
\label{zpwidth}
\end{eqnarray}
%-----
This is the expression for a partial width of the $Z$ boson studied at LEP1
as they are used in the {\tt ZEFIT} package for fits including the
$Z,Z'$ mixing. 
  
\nn
One now can study the expressions which result
in case of both small mixing and weak loop corrections:
%-----
\begin{eqnarray}
{\bar a}_f(1) \equiv a_{f}^{eff} \equiv (1-y_f) \sqrt{\rho_f} \; a_f
= \sqrt{\rho_f} (c_M \; a_f + s_M \; a_f').
\end{eqnarray}
%-----
A naive ansatz for the same effective coupling could be:
%-----
\begin{eqnarray}
a_{f}^{eff} = c_M (\sqrt{\rho_f} \; a_f) + s_M a_f'.
\end{eqnarray}
%-----
A direct comparison shows that one should not expect a more than tiny
difference. \\
For the resulting effective vector coupling, one gets similarly:
%----
\begin{eqnarray}
{\bar v}_f(1) = \rho_f (c_M a_f + s_M a'_f) [1-4\sin^2
\theta_W(1-x_f)\kappa_f]. 
\end{eqnarray}
%----
A naive ansatz could be here:
\begin{eqnarray}
v_{f}^{eff} = c_M [\sqrt{\rho_f} \; a_f( 1-4 \sin^2 \theta_W |Q_f| \kappa_f)]
+ s_M v'_f.
\end{eqnarray}
%%TR: until here the input done 10-7-1998.
%--------------------------------------------------------------
%\input{zeffsig}      % f.f. for cross sections    sect. 4
\section{Weak Form Factors of the Scattering Matrix Element
in Presence of $Z,Z'$ Mixing}
The Born matrix element for the scattering through the mass
eigenstate $Z_1$ (being observed at LEP1) is:
%----------------------------------
\begin{equation}
{\bf\cal M}_{1}(s,\cos \vartheta)  \sim \frac{1}{s-m_1^2} a_e(1)
a_f(1) 
\left[ \frac{G_{\mu}}{\sqrt{2}} M_Z^2\right] 
  \left[ \gamma_{\mu} \left( \frac{v_e(1)}{a_e(1)} + \gamma_5 \right)
       \right] \otimes
  \left[ \gamma^{\mu} \left( \frac{v_f(1)}{a_f(1)} + \gamma_5 \right)
      \right],
\label{mzp0}
\end{equation}
where $m_1^2$ denotes the complex mass parameter including finite width 
effects\footnote{
We do not discuss here problems connected with the definition of gauge 
boson masses in dependence on the handling of the energy dependence of 
the width.
}.
The following short notation is used:
\begin{equation}
A_{\gamma} \otimes B^{\gamma} = \left[ \bar u_e A_{\gamma} u_e \right]
                          \cdot \left[ \bar u_f B^{\gamma} u_f \right].
\end{equation}
A similar ansatz may be written in the t-channel. The matrix element
may be rewritten in terms of standard theory (unmixed) variables
$a_{e,f}, \sin^2 \theta_W$:
\begin{eqnarray}
&& {\bf\cal M}_{1}(s,\cos \vartheta)  \sim 
\nonumber  \\
&&\frac{1}{s-m_1^2} a_e a_f 
\left[ \frac{G_{\mu}}{\sqrt{2}} M_Z^2\right] 
%G_{\mu}
(1-y_e)
                             (1-y_f)
      [ L_{\mu} \otimes L^{\mu}
    - 4|Q_e |\sin^2 \theta_W (1-x_e) \gamma_{\mu} \otimes L^{\mu}
\nonumber  \\
&&    - 4|Q_f |\sin^2 \theta_W (1-x_f) L_{\mu} \otimes \gamma^{\mu}
%\nonumber  \\
    + 16 |Q_e Q_f | \sin^4 \theta_W (1-x_e)(1-x_f) \gamma_{\mu          }
              \otimes     \gamma^{\mu} ],
\label{mzfs}
\end{eqnarray}
\begin{eqnarray}
L_{\mu} = \gamma_{\mu}(1+\gamma_5 ),
\end{eqnarray}
where again we used that one can write
\begin{eqnarray}
\frac{v_f(1)}{a_f(1)} &=& 1 - 4|Q_f|\sin ^2 \theta_W  (1- x_f),
\\
a_f(1) &=&  (1-y_f) a_f,
\end{eqnarray}
with the same definitions of $x_f, y_f$ as in the case of the $Z$ width.
 
\nn
Without mixing, the weak loop corrections influence the matrix element as
follows:
%----
\begin{eqnarray}
&& {\bar{\cal M}}_{Z}(s,\cos \vartheta)  \sim 
\nonumber \\
&& \frac{1}{s-m_Z^2} 
\left[ \frac{G_{\mu}}{\sqrt{2}} M_Z^2\right] 
%G_{\mu} 
a_e a_f
          \rho(s,\cos \vartheta)
      [ L_{\mu} \otimes L^{\mu}
- 4|Q_e |\sin^2 \theta_W \kappa_e (s,\cos \vartheta ) \gamma_{\mu} \otimes
L^{\mu}
        \nonumber  \\
&& - 4|Q_f |\sin^2 \theta_W \kappa_f (s,\cos \vartheta ) L_{\mu} \otimes
\gamma^{\mu}
%         \nonumber  \\
    + 16 |Q_e Q_f | \sin^4 \theta_W \kappa_{ef}(s,\cos \vartheta) \gamma_{\mu}
\otimes
                                             \gamma^{\mu} ] .
\label{mzff}
\end{eqnarray}
%----
For massless fermions,
the four form factors $\rho, \kappa_e, \kappa_f, \kappa_{ef}$
are the most general ansatz for the weak radiative
corrections. In Born approximation, $\rho=\kappa=1$. \\
 
\nn
In case of $Z,Z'$ mixing, the matrix element may be written in a similar form:
%----
\begin{eqnarray}
&& {\bar{\cal M}}_{1}(s,\cos \vartheta)  \sim 
\nonumber \\
&& \frac{1}{s-m_1^2} 
\left[ \frac{G_{\mu}}{\sqrt{2}} M_1^2\right] 
%G_{\mu} 
a_e(1)
a_f(1)    \rho^M(s,\cos \vartheta)
\nonumber \\
&& \nobody    [ L_{\mu} \otimes L^{\mu}
- 4|Q_e |\sin^2 \theta_W \kappa_e^M (s,\cos \vartheta ) \gamma_{\mu} \otimes
L^{\mu}
  \nonumber  \\
&& -~ 4|Q_f |\sin^2 \theta_W \kappa_f^M (s,\cos \vartheta ) L_{\mu} \otimes
\gamma^{\mu}
%         \nonumber  \\
    + 16 |Q_e Q_f | \sin^4 \theta_W \kappa_{ef}^M(s,\cos \vartheta)
\gamma_{\mu}
\otimes
                                             \gamma^{\mu} ],
\label{mzff2}
\end{eqnarray}
%-----------
where it is:
\begin{eqnarray}
\rho^M &=& \rho_{mix}(1-y_e)(1-y_f) \rho,
\\
\kappa_{f}^M &=& (1-x_f) \kappa_{f},
%----------------------
\\
\kappa_{ef}^M &=& (1-x_e)(1-x_f) \kappa_{ef}.
\end{eqnarray}
%----------------
The matrix element may be rewritten in terms of effective weak
neutral vector and axial vector couplings:
\begin{eqnarray}
{\bar{\cal M}}_{1}(s,\cos \vartheta)  \sim \frac{1}{s-m_1^2} 
\left[ \frac{G_{\mu}}{\sqrt{2}} M_1^2\right] 
%G_{\mu}
      [ \bar a_e \bar a_f \gamma_{\mu} \gamma_5 \otimes
                                \gamma_{\mu} \gamma_5
      + \bar v_e \bar a_f  \gamma_{\mu} \otimes \gamma_{\mu} \gamma_5
      + \bar a_e \bar v_f  \gamma_{\mu} \gamma_5 \otimes \gamma^{\mu}
              \nonumber \\
      +        \bar v_{ef} \gamma_{\mu} \otimes \gamma^{\mu} ],
\end{eqnarray}
\begin{eqnarray}
\bar a_f(1) &=& \sqrt{\rho^M (s,\cos \vartheta)} \: I_3^L (f),
\label{axial2}
\\
\bar v_f(1) &=& \bar a_f(1) \left[ 1 - 4 |Q_f |\sin^2 \theta_W
                                        \kappa^M_f (s   ,\cos \vartheta)
                    \right],
\\
\bar v_{ef}(1) &=& \bar a_e(1) \bar v_f(1) + \bar v_e(1)
                                 \bar a_f(1) - \bar a_e(1) \bar a_f(1)
              \left[ 1 - 16 |Q_e Q_f |\sin^4 \theta_W \kappa^M_{ef} (s,\cos
\vartheta)
              \right].
\end{eqnarray}
%\nn
An equivalent notation is (with and without mixing):
\begin{eqnarray}
\bar a_f &=& \sqrt{\rho} a_f,
\label{baf}
\\
\bar v_f &=& \sqrt{\rho} \kappa_f v_f +  (1-\kappa_f) \bar a_f,
\label{bvf}
\\
\bar v_{ef} &=& \rho \kappa_{ef} v_e v_f + (\bar v_e - \sqrt{\rho}
\kappa_{ef} 
  v_e) \bar a_f + \bar a_e (\bar v_f - \sqrt{\rho} \kappa_{ef} v_f)
       - (1-\kappa_{ef}) \bar a_e \bar a_f.
\label{bvef}
\end{eqnarray}

\nn
Alternatively, one could define the axial vector couplings to be
unchanged by the radiative corrections. Then, the Fermi constant
absorbs the weak form factor $\rho(s,\cos \vartheta)$ and becomes dependent on
the process and its kinematics:
\begin{equation}
G_{\mu}  \rightarrow \bar G^M_{\mu} = G_{\mu} \rho^M(s,\cos \vartheta).
\label{gmu}
\end{equation}
The other form factors renormalise the weak mixing angle $\sin^2\theta_W
= 1-M_W^2/M_Z^2$:
\begin{equation}
\sin^2\theta_W  \rightarrow \left\{
                            \begin{array}{l} \sin^2 \theta_W \kappa^M_e(s,\cos
\vartheta) \\
                                          \sin^2 \theta_W \kappa^M_f(s,\cos
\vartheta) \\
                                 \sin^2 \theta_W \sqrt{\kappa^M_{ef}(s,\cos
\vartheta)}
                            \end{array}
                            \right. .
\label{swkap}
\end{equation}

\nn
Finally, one should mention that the mixing of gauge bosons
\z, \zp $\;$ as discussed here could be mimicked by a mixing of standard
fermions with exotic fermions \cite{673}, \cite{buchmuwi}\footnote
{In \cite{buchmuwi}, the common influence of exotic fermions and a
$Z'$ is discussed.} 
with quite similar influence on
(\ref{eq31},\ref{eq32}).
 
\nn
In order to simplify a comparison of the present approach to that of other
groups, we now give the explicit leading top quark mass $m_t$ dependence of
the form factors and $\Delta r$
\cite{Rbha,Rchj,djouadi,kniehl}
 in case of mixing:
%----
\begin{eqnarray}
\sin^2 \theta_W M_W^2 &=& \frac {\pi \alpha / (\sqrt{2} G_{\mu})}
                              {1 - \Delta r},
\\
%----
\sin^2 \theta_W &=& 1 - \frac {M_W^2} {M_Z^2}
= 1 - \frac {M_W^2} {M_1^2 \rho_{mix}} ,
\end{eqnarray}
%----
and $\rho_{mix}$ is defined in (\ref{rhomix}).
Further\footnote
{We follow the
conventions of {\tt ZFITTER}4 \cite{zfitter} with the following setting of
flags:
IAMT4=2, IQCD $\neq$ 0. The actual version of {\tt ZEFIT} is to be used
together with {\tt ZFITTER} v.3.05 until the official release of version 4.
The corresponding flag setting in version 3 is: IAMT4=1. The resummation
of the higher order QCD terms is not realized there.
},
%----
\begin{equation}
\Delta r = 1 + \Delta r^{rem} - (1 + \frac {\cos^2 \theta_W}{\sin^2 \theta_W}
\delta {\bar{\rho}} ) (1 - \Delta \alpha),
\end{equation}
%----
\begin{equation}
\delta {\bar{\rho}} = 3 {\cal T} \left[ 1 - (2 \pi^2 - 19) {\cal T}
- \frac {\alpha_s}{\pi} \frac{2}{3} (1+ \frac{\pi^2}{3}) \right],
\end{equation}
%----
\begin{equation}
{\cal T} = \frac{G_{\mu}}{\sqrt{2}} \frac{m_t^2}{8 \pi^2}.
\end{equation}
%----
Here,
$\alpha_s$ is the strong interaction coupling constant, and $\Delta r^{rem}$
contains the \oalf corrections  to $\Delta r$ without the contribution
${\cal T}$. The $\Delta \alpha$ contains the fermionic one-loop insertions to
$\alpha$.
For the cross section form factors, analogue formulae hold:
%----
\begin{eqnarray}
\rho^M &=&\rho_{mix}(1-y_e)(1-y_f)  \frac{1+\Delta \rho^{rem}_f}{1-\delta \bar\rho},
%----
\\
\kappa^M_f &=& (1-x_f) (1+\Delta \kappa^{rem}_f)
(1 + \frac {\cos^{2}\theta_{W}}  {\sin^{2}\theta_{W}} \delta \bar\rho),
%----
\\
\kappa^M_{ef} &=&(1-x_e) (1-x_f) (1+\Delta \kappa^{rem}_{ef})
(1 + \frac {\cos^{2}\theta_{W}}  {\sin^{2}\theta_{W}} \delta \bar\rho)^2.
\end{eqnarray}
%----
For the partial $Z$ widths,
%----
\begin{eqnarray}
\rho^M_f &=& \rho_{mix}(1-y_f)^2  \frac{1+\Delta \rho^{rem}_f}{1-\delta \bar\rho},
%----
\\
\kappa^M_f &=& (1-x_f) (1+\Delta \kappa^{rem}_f)
(1 + \frac {\cos^{2}\theta_{W}}  {\sin^{2}\theta_{W}} \delta \bar\rho).
\end{eqnarray}
%----
The cross section form factors and those of the partial $Z$ widths differ by
their remnant parts and, in case of the $\rho$, by their mixing factors.
At very high energy, the remnant parts of form factors become substantial
and both the approximate flavor independance, the similarity of width and
cross section form factors, and the factorization property of $\kappa_{ef}$
are lost.
 
\nn We remind the reader here that it was not our intention to discuss the
various
weak mixing angle definitions which allow an aproximate short hand notation
for partial $Z$ widths and cross sections.
We introduced the weak
corrections such that they are exact to weak one loop order
after introduction of the form factors. \\
A good approximation to the so-called effective weak mixing angle which often
is used for a description of LEP1 data is
%----
\begin{equation}
\sin^2\theta_{W,eff} = \kappa \; \sin^2 \theta_W,
\end{equation}
%----
where one can take any one of the (real part of)
form factors $\kappa_f$, calculated at
$s=M_Z^2, \cos\vartheta=0$
or the corresponding form factor from a partial width, e.g. $\Gamma_e$.
For some details see \cite{15} and for a
comparative discussion \cite{sommi}.
 
\nn
At the end of this section, we should mention that all the above derivations
are valid for Bhabha scattering (\ref{eeee}), too. This is due to the fact
that the discussion has been 
based on the matrix elements ${\bf\cal M}$ before building cross
sections out of them.

%--------------------------------------------------------------
%\input{zesigma}      % improved Born cross section  sect. 5
\section{Improved Born Approximation in Presence of Gauge Boson
Mixing}
The matrix element ${\bf\cal M}_{Z'}$ with $Z'$ exchange is in case of
mixing:
\begin{eqnarray}
{\bf\cal M}_{2}(s,\cos \vartheta)  &\sim& \frac{{g'}^2}{s-m_2^2}
    [ \bar a_e(2) \bar a_f(2) \gamma_{\mu} \gamma_5 \otimes
                                \gamma_{\mu} \gamma_5
      + \bar v_e(2) \bar a_f(2)  \gamma_{\mu} \otimes \gamma_{\mu} \gamma_5
\nonumber \\
   && +~ \bar a_e(2) \bar v_f(2)  \gamma_{\mu} \gamma_5 \otimes \gamma^{\mu}
      +  \bar v_{ef}(2) \gamma_{\mu} \otimes \gamma^{\mu} ],
\end{eqnarray}
where $\bar a_f(2), \bar v_f(2) $ are the renormalized vector and axial vector
couplings of the $Z_2$.\\
As long as we search for a \zp $\;$ well below the production threshold, we
can safely neglect here all radiative corrections to this amplitude,
\begin{equation}
\bar a_f(2) = a_f(2), \hspace{1cm} \bar v_f(2) = v_f(2),
\hspace{1cm} \bar v_{ef}(2) = v_e(2) v_f(2).
\end{equation}
%
%\vspace{1cm}
 
\nn
In sum, our discussion leads to the following net matrix element,
where we also add up the photon exchange diagram with running QED coupling:
\begin{equation}
{\bf\cal M} = {\bf\cal M}_{\gamma} + {\bf\cal M}_{1} +
                {\bf\cal M}_{2}.
\end{equation}

\nn
Let us now have a look at the squared matrix elements.
In case of massless fermion production,
four different combinations of coupling constants may occur in reactions
(\ref{ee},\ref{ep}) from the interference of the vector bosons $m$ and
$n$:
\begin{eqnarray}
\bar C_1(m,n) &=
 & \bar a_e(m) \bar a^*_e(n) \bar a_f(m) \bar a^*_f(n) +
  \bar a_e(m) \bar a^*_e(n) \bar v_f(m) \bar v^*_f(n) 
\\ \nonumber
 && +~ \bar v_e(m) \bar v^*_e(n) \bar a_f(m) \bar a^*_f(n) +
  \bar v_{ef}(m)          \bar v^*_{ef}(n),
\\
\bar C_2(m,n) &=
&  \bar a_e(m) \bar v^*_e(n) \bar v_{f}(m) \bar a^*_f(n) +
  \bar v_e(m) \bar a^*_e(n) \bar a_f(m)   \bar v^*_f(n) 
\\ \nonumber
 && +~\bar a_e(m) \bar v^*_{ef}(n) \bar a_f(m)              +
  \bar a^*_e(n) \bar v_{ef}(m)            \bar a^*_f(n),
\\
\bar C_3(m,n) &=
&  \bar a_e(m) \bar v^*_e(n) \bar a_f(m) \bar a^*_f(n) +
  \bar v_e(m) \bar a^*_e(n) \bar a_f(m) \bar a^*_f(n) 
\\ \nonumber
&& +~ \bar a_e(m) \bar v^*_{ef}(n) \bar v_f(m) +
  \bar a^*_e(n) \bar v_{ef}(m)          \bar v^*_f(n),
%            1
\\
\bar C_4(m,n) &=
&  \bar a_e(m)   \bar a^*_e(n)    \bar a_f(m) \bar v^*_f(n) +
  \bar a_e(m)   \bar a^*_e(n)    \bar v_f(m) \bar a^*_f(n) \\ \nonumber
 && +~\bar v_e(m)   \bar v^*_{ef}(n) \bar a_f(m) +
  \bar v^*_e(n) \bar v_{ef}(m)               \bar a^*_f(n).
\end{eqnarray}
The starred couplings of vector boson $n$ and the corresponding
propagators are complex conjugated;
a procedure which is necessary only in the s-channel.

Now we have all formulae needed to write down the
improved Born cross section.
For initial state radiation, they read\footnote{
While for final state radiation the $s'$ is replaced by $s$, the
$s$-dependence is more complicated for the initial-final state
interference \cite{zfitter,351}.
We further remind the reader that {\tt ZFITTER} doesn't return
differential cross sections at all; (\ref{dsdc})
is shown for illustrational purposes. 
}:
%-------------------------------------------
\begin{eqnarray}
\sigma^0_{T} &=& \frac{\pi\alpha^2}{2s'} \Re e\sum_{m,n=0}^N
\chi_m(s')\chi_n^*(s')
\Bigl[ \bar C_1(m,n) L_1 H_1 
\\ \nonumber
&& +~ \bar C_3(m,n) L_2 H_1 +
 \bar C_4(m,n) L_1 H_2 +\bar C_2(m,n) L_2 H_2 \Bigr],
\label{sigmas}
\\
\sigma^0_{FB} &=& \frac{\pi\alpha^2}{2s'} \Re e\sum_{m,n=0}^N
\chi_m(s')\chi_n^*(s')
\Bigl[ \bar C_2(m,n) L_1 H_1 
\\ \nonumber &&
+~\bar C_4(m,n) L_2 H_1 +
 \bar C_3(m,n) L_1 H_2 +\bar C_1(m,n) L_2 H_2 \Bigr],
\\
 \chi_n(s)&=&\frac{g_n^2}{4\pi\alpha}\frac{s}{s-m_n^2},
\label{sigmafb}
%-------------------
\\
\frac{d\sigma}{d\cos\vartheta} &=&  (1+\cos^2\vartheta) \sigma^0_{T}
                              +  (2 \cos \vartheta) \sigma^0_{FB}.
\label{dsdc}
\end{eqnarray}
The quantities $L_1,\ L_2,\ H_1$ and $H_2$ are combinations of the
polarizations of the beams ($e^-, e^+$) and of the helicities of the
final fermions: 
%-------------------------------------------
\begin{equation}
L_1=1-\lambda_+\lambda_-,\hspace{0.5cm}
L_2=\lambda_+ -\lambda_-,\hspace{0.5cm}
H_1=\frac{1}{4} (1-h_+h_-),\hspace{0.5cm}
H_2=\frac{1}{4} (h_+ -h_-),
\label{heli}
\end{equation}
%--------------------
with $\lambda_-$ and $\lambda_+$ ($h_-$ and $h_+$) being the polarizations
of electron and positron (fermion and antifermion).
%In {\tt ZFITTER} the corresponding mass terms for the exchange of vector
%bosons $m$ and $n$ in the Born have been added.
The masses in the propagator $\chi_n(s)$ are:
\begin{eqnarray*}
 m_0^2= & 0 & \hspace{1.cm} \gamma\nonumber\\
 m_1^2= & M_1^2-iM_1\Gamma_1 & \hspace{1.cm} Z\nonumber\\
 m_2^2= & M_{2}^2-iM_{2}\Gamma_{2} & \hspace{1.cm} Z' .
\label{m012}
\end{eqnarray*}
%-------------------------------------------
In case of a (\z,\z) Born cross section with a polarization of the
electron beam, the following combinations correspond
to the well-known coupling factors:
%--------------
\begin{equation}
C_1(Z,Z) + \lambda_+ C_3(Z,Z) = (a_e^2 + v_e^2 + 2 \lambda_+ a_e v_e)
                              (a_f^2 + v_f^2),
\end{equation}
\begin{equation}
C_2(Z,Z) + \lambda_+ C_4(Z,Z) = [2 a_e v_e + \lambda_+ (a_e^2 + v_e^2)]
                                (2 a_f v_f).
\end{equation}

With the above definitions, we have all the necessary prerogatives
to calculate cross sections with both weak loop effects
and \zp $\;$ exchange. The resulting Born formulae can be used
as input for a QED calculation.\\
 
\nn
The above expressions for $\sigma^0_{T}$ and $\sigma^0_{FB}$ are used
in the redefinitions to be performed in the subroutine {\tt BORN} of
{\tt ZFITTER}.
%--------------------------------------------------------------
%\input{zefit3}       % program description   sect. 6
\section{Structure of {\tt ZEFIT}}
The package {\tt ZEFIT} should be run together with {\tt ZFITTER}.
For execution
it has to be loaded before the package {\tt ZFITTER} since it contains
subroutines which originate from {\tt ZFITTER} but are modified for the
description of the $Z'$.
\\ Searching for a $Z'$, we assume that the $Z$ boson interactions
are correctly described by the Standard Model.
Then there are two ways to use the package. {\em Either} one performs a
model-independent fit to data with {\tt ZFITTER} and searches with the
result - partial widths or effective couplings - for a $Z'$, using
basically the {\tt ZWRATE} or {\tt ROKANC} subroutines of 
{\tt DIZET}\footnote{
{\tt ZWRATE} calls {\tt ROKAPP} to calculate the weak form factors of
the widths.},
{\em or} 
one tries an immediate fit to the cross sections, using basically
subroutine {\tt ZCUT} calling {\tt EWINIT} of {\tt ZFITTER}\footnote{
{\tt EWINIT} uses the subroutine {\tt ROKAP} in {\tt ZFITTER} v.3.05 
({\tt ROKANC} in  {\tt ZFITTER} v.4) to calculate the form factors of
the cross sections.}. 
The common
package 
{\tt ZEFIT}$\oplus${\tt ZFITTER} is prepared for both applications.
The second
approach, within the (possibly extended) Standard Model terminology,
is technically more involved.
So, we will use for the present purpose of demonstration
only 
the corresponding interface and branch with subroutine {\tt ZANALY}
of {\tt ZFITTER} v.3.05. 
\\
\\ \underline{\bf Input arguments:}
\\ \begin{description}
    \item[$\bf AMZ$]  is the mass of the $Z$ boson,
    \item[$\bf AMH$]  is the mass of the Higgs boson in GeV,
    \item[$\bf AMT$]  is the top mass in GeV,
    \item[$\bf INDF$] selects the final state fermion type,
    see also the table in the test example in Appendix~B,
 
    \item[$\bf SQS$] is the centre of mass energy in GeV,
    \item[${\bf IFAST} = 1$] allows a fast calculation without
    geometrical and kinematical cuts. It must be ${\bf IFAST}=0$ if
    cuts are required,
    \item[${\bf IRCUT} = 0,1$] chooses between an acollinearity cut
    and a cut on the photon energy.
   \end{description}
 New additional input parameters coming from a $Z'$:
  \begin{description}
    \item[$\bf AMZE$] is the mass of the $Z'$ in GeV,
    \item[$\bf ZMIX$] is the mixing angle between $Z$ and $Z'$,
    \item[${\bf IZE} = 0,1,2$] chooses the model (Standard Model,
    $E_6$ model, left-right model).
\end{description}
As examples,
we have foreseen a $Z'$ coming from an $E_6$-GUT and from a left-right
symmetric model.
The model chosen must be specified by the flag  $\bf IZE$.
The Standard Model is realised for $\bf{ IZE} = 0$, the $E_6$ model
is chosen with $\bf{ IZE} = 1$ and the LR model with $\bf{ IZE} = 2$.
Inside the $E_6$ model the parameter $\bf TETAE6$
must be set as the
mixing angle of the two extra $Z$ generators.
For the left-right symmetric model the parameter $\bf ANGLR$
has to be defined. ${\bf ANGLR}$ is limited,
i.e.
$\sqrt{\frac{2}{3}} \leq {\bf ANGLR} \leq \sqrt{2}$.
The unit for the parameters ${\bf TETAE6}$ and ${\bf ANGLR}$ in {\tt
ZEFIT} is radian. 
\\
\\ \underline{\bf Additional flags of {\tt ZFITTER}}
The flags   for weak loop and QED
corrections are     set in the subroutine {\tt ZINITF} \cite{zfitter} :
\begin{description}
 \item[$\bf IWEAK$] = 0 or 1, switches the $O(\alpha)$ weak loops.
 \item[$\bf IHVP$] = 1, 2, 3 characterises  the vacuum polarization
  parametrization. The best choice is {\bf IHVP} = 3.
 \item[$\bf IQCD$] = 0, 1, 2, 3, 4 gives the QCD
 corrections to the vector
  boson self energies in ${\Delta}r$, the widths and the cross section.
  {\bf IQCD} = 3 is recommended for LEP applications.
 \item[$\bf IAMT4$] = 0, 1; the leading two loop effects of the type
  $O(\alpha^{2}m_{t}^4)$ can be included.
 \item[$\bf IBOX$] = 0, 1; WW and ZZ box corrections may be taken into
  account.
 \item[$\bf IFINAL$] = 0, 1 chooses between the approximate final
  state correction by the factor
 \ $[1 + 3\pi Q^2_f/(4\pi)]$
  and the correct $O(\alpha)$ final state correction including soft
photon 
  exponentiation.
 \item[$\bf INTERF$] = 0, 1 switches the $O(\alpha)$ initial -
  final state interference.
 \item[$\bf IPHOT2$] = 0, 1; second order QED leading logs may be taken
  into account.
\end{description}
Additionally to these flags of {\tt ZFITTER}, we introduced
\\ $\bf IALFRUN$ = 0,1; which allows an independent switching off
and on of running $\alpha_{QED}$.
\vspace*{.5cm}

\nn
 \underline{\bf Output parameters:}
\\
 \begin{description}
  \item[$\bf SBORN$] - Born cross section in nb.
  \item[$\bf STOT$] - total cross section in nb.
  \item[$\bf ABORN$] - Born forward-backward asymmetry.
  \item[$\bf ATOT$] - forward-backward asymmetry.
 \end{description}
In fig. 1 the block diagram of  {\tt ZEFIT} is shown.
The subroutines in the dashed boxes remained unchanged.
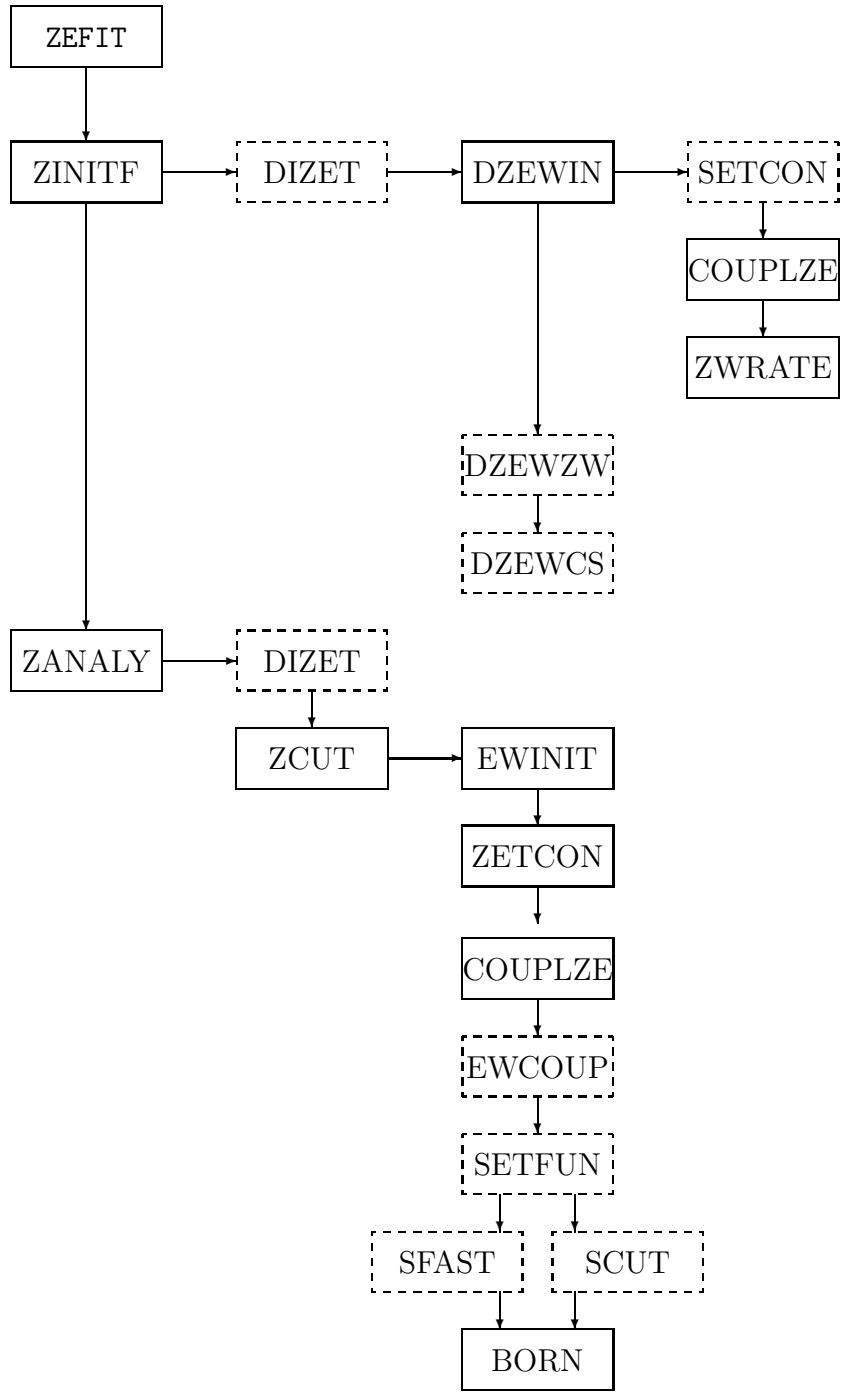
\begin{figure}
\setlength{\unitlength}{1mm}
\begin{picture}(140,190)(-12.5,0)
\put(20,190){\framebox(20,8){{\tt ZEFIT}}}
\put(30,190){\vector(0,-1){10}}
\put(20,172){\framebox(20,8){ZINITF}}
\put(30,172){\vector(0,-1){57}}
\put(40,176){\vector(1,0){10}}
\put(50,172){\dashbox(20,8){DIZET}}
\put(70,176){\vector(1,0){10}}
\put(80,172){\framebox(20,8){DZEWIN}}
\put(100,176){\vector(1,0){10}}
\put(110,172){\dashbox{1}(20,8){SETCON}}
\put(120,172){\vector(0,-1){ 5}}
\put(110,159){\framebox(20,8){COUPLZE}}
\put(120,159){\vector(0,-1){ 5}}
\put(110,146){\framebox(20,8){ZWRATE}}
\put(90,172){\vector(0,-1){31}}
\put(80,133){\dashbox{1}(20,8){DZEWZW}}
\put(90,133){\vector(0,-1){ 5}}
\put(80,120){\dashbox{1}(20,8){DZEWCS}}
\put(20,107){\framebox(20,8){ZANALY}}
\put(40,111){\vector(1,0){10}}
\put(50,107){\dashbox(20,8){DIZET}}
\put(60,107){\vector(0,-1){5}}
\put(50,94){\framebox(20,8){ZCUT}}
\put(70,98){\vector(1,0){10}}
\put(80,94){\framebox(20,8){EWINIT}}
\put(90,94){\vector(0,-1){5}}
\put(80,81){\framebox(20,8){ZETCON}}
\put(90,81){\vector(0,-1){5}}
\put(80,66){\framebox(20,8){COUPLZE}}
\put(90,66){\vector(0,-1){5}}
\put(80,53){\dashbox{1}(20,8){EWCOUP}}
\put(90,53){\vector(0,-1){5}}
\put(80,40){\dashbox{1}(20,8){SETFUN}}
\put(95,40){\vector(0,-1){5}}
\put(92,27){\dashbox{1}(20,8){SCUT}}
\put(85,40){\vector(0,-1){5}}
\put(68,27){\dashbox{1}(20,8){SFAST}}
\put(85,27){\vector(0,-1){5}}
\put(95,27){\vector(0,-1){5}}
\put(80,14){\framebox(20,8){BORN}}
\end{picture}
\caption{ {\bf Structur of the package {\tt ZEFIT}.}
Subroutines in solid boxes are rewritten and replace the corresponding
subroutines of {\tt ZFITTER} in case of $Z, Z'$ mixing. The subroutines in dashed
boxes remain unchanged. They are shown here in order to illustrate the
interplay of {\tt ZEFIT} with {\tt ZFITTER}.}
\label{fi:eins}
\end{figure}
\clearpage
\newpage
\nn
{\Large\bf Appendix A\\
           Additional parameters in theories with extra $Z$ bosons}
\vspace{0.5cm} \\
\setcounter{equation}{0}
\renewcommand{\theequation}{A.\arabic{equation}}
%\renewcommand{\theequation}{\thesection.\arabic{equation}}
%Theories involving a gauge group larger than SU(5) typically predict the
%existence of extra Z bosons \cite{verz},\cite{673}.
The couplings of $Z_1$ and $Z_2$ to fermions are in general a result of a
mixing between the Standard Model $Z$ and the $Z'$
as introduced in (\ref{zzmix}):
%-----------------------------------------
\begin{eqnarray}
v_f(1)=   \cos \theta_M\; v_f + \; \sin\theta_M\;  \frac{g'}{g} v'_f,
\nonumber\\ 
a_f(1)=   \cos \theta_M\; a_f + \; \sin\theta_M\;  \frac{g'}{g} a'_f,
\nonumber \\ 
v_f(2)=   \cos \theta_M\; v'_f - \; \sin\theta_M\; \frac{g}{g'} v_f,
\nonumber\\ 
a_f(2)=   \cos \theta_M\; a'_f - \; \sin\theta_M\; \frac{g}{g'} a_f.
\end{eqnarray}
%-------------------------------------------
The couplings of the extra Z boson to fermions $a_f'$ and $v_f'$
depend on the particular model. In {\tt ZEFIT} we have implemented extra
$Z$ bosons coming from an $E_6$ GUT or from a left-right model.
 
\nn
In the $E_6$ GUT the $E_6$ group \cite{19} is assumed
to be broken to the standard model
group structure in the following way \cite{20}:
\begin{eqnarray}
E_6 \longrightarrow SO(10) \times U(1)_\psi \longrightarrow
SU(5) \times U(1)_\chi \times U(1)_\psi \longrightarrow \nonumber \\
\longrightarrow
SU(3)_c \times SU(2)_L \times U(1)_Y \times U(1)_\chi \times U(1)_\psi .
%\label{eq33}
\end{eqnarray}
We further assume that the following linear combination of
the two extra $Z$ bosons $Z_\chi$ of $U(1)_\chi$ and $Z_\psi$ of $U(1)_\psi$ is
light and can be detected at future colliders \cite{21}:
\begin{equation}
Z' =\cos \theta_E\;Z_\chi + \sin \theta_E\;Z_\psi.
\label{eq34}
\end{equation}
The couplings between the $Z'$ and fermion $f$ are \cite{21p}:
\begin{equation}\begin{array}{lr lr}
a'_\nu = & \frac{3}{2}Q_\chi\cos\theta_E\;+\;\frac{1}{2}Q_\Psi\sin\theta,
\hspace{1cm}
v'_\nu = & \frac{3}{2}Q_\chi\cos\theta_E\;+\;\frac{1}{2}Q_\Psi\sin\theta,
\vspace{.3cm}
\\
a'_e  = & Q_\chi\cos\theta_E\;+\;Q_\Psi\sin\theta,
\hspace{1cm}
v'_e   = & 2 Q_\chi\cos\theta_E,
\vspace{.3cm}
\\
a'_u  = & -Q_\chi\cos\theta_E\;+\;Q_\Psi\sin\theta,
\hspace{1cm}
v'_u  = & 0,
\vspace{.3cm}
\\
a'_d  = & Q_\chi\cos\theta_E\;+\;Q_\Psi\sin\theta,
\hspace{1cm}
v'_d  = & -2 Q_\chi\cos\theta_E,
\\
\label{eq35}     \end{array}
\end{equation}
\begin{equation}
Q_\chi  = \frac{1}{\sqrt{10}}, \hspace{1cm} Q_\psi  = \frac{1}{\sqrt{6}}.
\end{equation}
%--------------------------
   The $Z'$ coupling constant $g'$  may be determined by the assumption
that the renormalisation group evolution of $g'$  is the same as that
of $g$ \cite{23},
%--------------------------
\begin{equation}
g' = \sqrt{\frac{5}{3}} \; \sin\theta_W\;g .
\label{eq36}
\end{equation}
%--------------------------
Of particular interest for applications is the completely specified case
$Z' \equiv Z_\eta = \sqrt{3/8}\;Z_\chi - \sqrt{5/8}\;Z_\psi$ as suggested by
superstring theories \cite{5,24}.
The most general $Z'$ in a $E_6$ GUT can therefore be described by three
additional
parameters: $\theta_M=$ZMIX, $\theta_E=$TETAE6 and $M(Z_2)=$AMZE.\\
Another origin of a $Z'$ could be a left-right symmetric model \cite{LRM}.
As in the $E_6$ case, we have $\theta_M=$ZMIX and $M(Z_2)=$AMZE
and one additional parameter $\alpha=$ANGLR.
In this model we have the following couplings of the $Z'$ to the
fermions:
\begin{equation}\begin{array}{rl rl}
a'_{\nu} = &
\displaystyle{
\frac{1}{2\alpha}},
& %\hspace{1cm}
v'_{\nu} = & a'_{\nu},
\vspace{.4cm}
\\
a'_e  = &
\displaystyle{
\frac{1}{2}\alpha },
& %\hspace{1cm}
v'_e  = &
\displaystyle{
\frac{1}{\alpha}  } \;-\; a'_e,
\vspace{.4cm}
\\
a'_u  = & -
\displaystyle{
\frac{1}{2}}\alpha,
&  \hspace{1cm}
v'_u  = & -
\displaystyle{
\frac{1}{3\alpha}  }  \;-\; a'_u,
\vspace{.4cm}
\\
a'_d  = &
\displaystyle{
\frac{1}{2}  } \alpha,
&  %\hspace{1cm}
v'_d  = & -
\displaystyle{
\frac{1}{3\alpha}   }  \;-\; a'_d.
\end{array}
\label{eqa35}
\end{equation}
%--------------------------
The relation between $g$ and $g'$ is:
%--------------------------
\begin{equation}
g' = \sin\theta_W\;g .
\label{eqa36}
\end{equation}
%--------------------------
%
Concerning the width of the $Z'$, we will assume that it can decay
only into particles of the known three fermion generations
including the top-quark.
%**********************************************
 
\nn
Of course, one also can make no model assumptions about the $Z'$
at all. Then, one has to specify arbitrarily
all needed couplings of the $Z'$
to fermions and the width of the $Z'$.
In addition, the $Z, Z'$ mixing angle and the $Z'$ mass have
to be given.
%--------------------------------------------------------------
%\input{zeappout}    % app. B     printout of test
\newpage
\nn
{\Large \bf Appendix B \\
            The Main Routine of the Test Program}
\begin{verbatim}
       PROGRAM ZEFIT
c
c
**********************************************************************
**                                                                  **
**     ZEFIT  -  a package for extra Z searches at LEP              **
**                                                                  **
**                                                                  **
**  A package [1] ZPRIME physics in presence of QED and weak        **
**  loop corrections as realised  in ZFITTER [2]                    **
**  A possible mixing of Z and Z' is taken into account.            **
**                                                                  **
**  Version:       2.00                                             **
**  Release:       November 25,1991                                 **
**  Authors:       A. Leike, S. Riemann, T. Riemann                 **
**                 Inst. for High Energy Physics, Zeuthen           **
**  Contact in L3: S. Riemann, riemanns@cernvm                      **
**                                                                  **
**  [1]  A.Leike, S.Riemann, T. Riemann:                            **
**       Z.Z' Mixing                                                **
**       in Presence of Standard Weak Loop Corrections,             **
**       Munich Univ. prepr. LMU-91/06 (Dec 1991)                   **
**  [2]  D.Bardin et.al.:                                           **
**       A Users Guide to ZFITTER: An Analytical Program for        **
**       Fermion Pair Production in e^+ e^- Annihilation,           **
**       in preparation.                                            **
**                                                                  **
**********************************************************************
*
       IMPLICIT REAL*8(A-H,O-W,Z)
       IMPLICIT COMPLEX * 16 (X,Y)
       REAL*4 TIME1,TIME2
       PARAMETER (NSARR=10,NMODEL=1)
*
*  NSARR  - NUMBER OF SQRTS POINTS TO BE CALCULATED
*  NMODEL - NUMBER OF POINTS SPECIFIED WITHIN AN EXTRA Z MODEL
*
      COMMON/ROVEFZ/ARROFZ(0:10),ARVEFZ(0:10)
      COMMON/FRINIT/ NPAR(30),ZPAR(30)
      COMMON/ZE/TETAMD,AMZE,GAMZP,ZMIX,CHIE
     +,         VEEZE,VEFZE,AEFZE,VQEZE(6),VQFZE(6)
     +,         XVEEZE,XVEFZE,XAEFZE,XVEFGZ,XAEFGZ
     +,         XVQEZE(6),XVQFZE(6),XVQFGZ(6)
       COMMON /IZE/ IZE
*  ARROFZ(0:9) - CALCULATED EFFECTIVE RO'S FOR EACH CHANNEL
*  ARVEVZ(0:9) - CALCULATED EFFECTIVE VECTOR COUPLINGS FOR EACH CHANNEL
*  ARROFZ(0:10) AND ARVEFZ(0:10) ARE UNDEFINED AND NOT USED
*  ARVEVZ      = 1 - 4 * ABS(QF) * SINTW**2 ( SINTW**2 IS EFFECTIVE )
      DIMENSION WIDTHS(0:10)
      DIMENSION TETAE6(NMODEL),ANGLR(NMODEL)
      DIMENSION SARR(NSARR)
* ARRAY OF CMS ENERGIES
      DATA SARR /87.d0,88.D0,89.0D0,90.d0,91.d0,91.17d0,92.d0,93.d0
     +,          94.d0,95.d0/
**
* CHOICE OF E_6 ZPRIME MODEL (TETA6 in RADIAN)
C     DATA TETAE6 /0.D0, 1.5708D0, -.91174D0/
      DATA TETAE6 /0.D0/
* CHOICE OF LEFT-RIGHT EXTRA Z MODEL (ANGLR in RADIAN)
      DATA ANGLR /0.81649658D0, 1.D0, 1.4142135D0/
C                   sqrt(2/3)          sqrt(2)
**
* INPUT FOR INITIALIZATION SUBROUTINE ZINITF:
* Z-MASS (ALL INPUT MASSES IN GEV)
      AMZ=91.180D0
      AMZLEP=AMZ
* TOP-MASS
      AMT=150D0
* HIGGS-MASS
      AMH=3D2
* ZPRIME MASS (NOT RELEVANT IF IZE = 0)
      AMZE= 5000.D0
* TWO QCD-CORRECTION FACTORS (THE SECOND ONE FOR B-BBAR CHANNEL)
      QCDCOR=1.040D0
      QCDCOB=1.045D0
*
* IZE = 1,2 (0) EXTRA Z CONTRIBUTIONS ARE (NOT) INCLUDED
*     = 0   STANDARD MODEL
*     = 1   E_6 MODEL (TETAE6 MUST BE SPECIFIED)
*     = 2   LEFT-RIGHT MODEL (ANGLR MUST BE SPECIFIED)
      IZE=1
* ZMIX - Z, ZPRIME MIXING ANGLE
      ZMIX = 0.01D0
* IF IFAST=1       WITHOUT CUTS
* IF IFAST=0       CUTS ARE POSSIBLE
      IFAST=1
* PLEASE, LOOK INTO ZINITF IN ORDER TO INITIALIZE ALL FLAGS AND 
* PARAMETERS WHICH COULD BE (AND SHOULD BE) INITIALIZED FOR A SPECIFIC 
* TASK.
*
      IF(IZE.EQ.0)  THEN
        INMOD=1
      ELSE
        INMOD=NMODEL
      ENDI
* SPECIFY THE MODEL DEPENDENT ANGLE
      DO 20  IZP=1,INMOD
* EXTRA Z FROM E_6 MODEL
      IF(IZE.EQ.1) TETAMD=TETAE6(IZP)
* EXTRA Z FROM LEFT-RIGHT MODEL
      IF(IZE.EQ.2) THEN
        ANGMIN = DSQRT(2.D0/3.D0)-1.D-5
        ANGMAX = DSQRT(2.D0)+1.D-5
        TETAMD=ANGLR(IZP)
        IF(TETAMD.LT.ANGMIN .OR. TETAMD.GT.ANGMAX) THEN
          PRINT*,' LEFT-RIGHT PARAMETER OUT OF RANGE'
          STOP
        ENDIF
      ENDIF
*
      CALL ZINITF(AMZ,AMT,AMH,QCDCOR,QCDCOB,SW2,WIDTHS)
* OUTPUT OF ZINITF:
* SW2 - CALCULATED QUANTITY = 1-AMW**2/AMZ**2
* WIDTHS(INDF) - CALCULATED PARTIAL CHANNEL Z-WITDHS (IN MEV)
*                FOR 0<INDF<9 AS DESCRIBED ABOVE
*                FOR   INDF=10 IT CONTAINS CALCULATED TOTAL Z-WIDTH
*
      IF(IZE.EQ.0)PRINT'(/,'' STANDARD MODEL, NO EXTRA Z EXTENSION'',/)'
      IF(IZE.EQ.1)PRINT'(/,'' EXTRA Z FROM E6 GUT'',/)
      IF(IZE.EQ.2)PRINT'(/,'' EXTRA Z FROM LEFT-RIGHT MODEL'',/)'
      PRINT 1000,AMZLEP,AMT,AMH
      IF(IZE.EQ.1)  PRINT 1100,AMZE,ZMIX,TETAMD
      IF(IZE.EQ.2)  PRINT 1200,AMZE,ZMIX,TETAMD
      PRINT 1300,QCDCOR,QCDCOB,SW2
      PRINT 1500,WIDTHS
      PRINT 1550,NPAR(1),NPAR(2),NPAR( 3),NPAR( 4)
     &,          NPAR(8),NPAR(9),NPAR(10),NPAR(16)
*
* GEOMETRICAL CUTS OVER THE ANGLE BETWEEN E+ AND OUTGOING ANTI-FERMION
      ANG1=140D0
      ANG2=40D0
      IF (IFAST .NE. 1) PRINT 1700,ANG2,ANG1
*
*  KINEMATICAL CUTS:
*  IF IRCUT=1 THEN SPRIME (IN GEV**2)
*  IF IRCUT=0 THEN ACOL+EMIN (IN DEGREES AND GEV, RESPECTIVELY)
*
      IRCUT = 0
      SPRIME= 1D-6
      ACOL  = 25.D0
      EMIN  = .0D0
      IF(IFAST.EQ.0) THEN
         IF(IRCUT.EQ.1) THEN
            PRINT 1800,SPRIME
         ELSE
            PRINT 1900,ACOL,EMIN
         ENDIF
      ELSEIF(IFAST.EQ.1) THEN
         PRINT 1600
      ELSE
         IFAST = 1
         PRINT 1600
      ENDIF
*
      DO 11 INDF=2,2
*  INDF=0 FOR NEUTRINO
*      =1 FOR ELECTRON
*      =2 FOR MUON
*      =3 FOR TAU
*      =4 FOR UP
*      =5 FOR DOWN
*      =6 FOR CHARM
*      =7 FOR STRANGE
*      =8 FOR TOP ( RETURNS ZERO )
*      =9 FOR BOTTOM
*     =10 FOR HADRONS
      PRINT 2000,INDF
      GAMEE=WIDTHS(1)/1D3
      GAMZ=WIDTHS(10)/1D3
      IF(INDF.NE.10) THEN
         GAMFI=WIDTHS(INDF)/1D3
      ELSE
         GAMFI=0D0
         DO 4 IH=4,9
 4       GAMFI=GAMFI+WIDTHS(IH)/1D3
      ENDIF
      IF(INDF .NE. 10) PRINT 2500
      IF(INDF .EQ. 10) PRINT 2600
* DO LOOP OVER THE SQRT(S)=SQS
      DO 10 IS=1,NSARR
      SQS=SARR(IS)
*
* NOW EVERYTHING WILL BE CALCULATED IN THE FRAMEWORK OF THE STANDARD
* MODEL BY ZANALY- SUBROUTINE.
*
      IF(INDF.NE.10) THEN
         CALL ZANALY(IFAST,INDF,SQS,IRCUT,SPRIME,ACOL,EMIN,ANG1,ANG2,
     &               AMZ,AMT,AMH,QCDCOR,QCDCOB,SBORN,STOT,ABORN,ATOT)
         CS =STOT
         AFB=ATOT
         IF(INDF.EQ.0) THEN
            PRINT 3000,SQS,CS
         ELSE
            PRINT 3100,SQS,CS,AFB
         ENDIF
      ELSE
         CS =0D0
         DO 53 INDFH=4,9
         CALL ZANALY(IFAST,INDFH,SQS,IRCUT,SPRIME,ACOL,EMIN,ANG1,ANG2,
     &               AMZ,AMT,AMH,QCDCOR,QCDCOB,SBORN,STOT,ABORN,ATOT)
         CS =CS +STOT
 53      CONTINUE
         PRINT 3000,SQS,CS
      ENDIF
 10   CONTINUE
 11   CONTINUE
 20   CONTINUE
1000 FORMAT(1X,/1X,'INPUT :  MZ    = ',F6.3,' GEV'
    &     ,'    MT    = ',F6.2,' GEV'
    &,     '    MH  = ',F8.2,' GEV'
1100 FORMAT ('          AMZE  = ',F6.1,' GeV'
    &,             '    ZMIX  = ',F6.4,'     TETAE6 = ',F8.4,/)
1200 FORMAT ('          AMZE  = ',F6.1,' GeV'
    &,             '    ZMIX  = ',F6.4,'      ANGLR = ',F8.4,/)
1300 FORMAT (' QCD CORRECTION FACTORS = ',2(F7.5,2X),/
    &       ,' OUTPUT:    SW2         = ',F5.4)
1500 FORMAT (1X,'PARTIAL AND TOTAL Z-WIDTHS IN MEV',/
    &,5x,'nu,nubar  =',F7.1,8x,'e+,e-  =',F7.1,6x,'mu,mubar =',F7.1
    &,/,5x,'tau+,tau- =',F7.1,8x,'u,ubar =',F7.1,8x,'d,dbar =',F7.1
    &,/,5x,'c,cbar    =',F7.1,8x,'s,sbar =',F7.1
    &,/,5x,'t,tbar    =',F7.1,8x,'b,bbar =',F7.1,8x,'TOTAL  =',F7.1
    &,/)
1550 FORMAT (/,' FLAGS IN INITIALISATION ROUTINE ZINITF:',
    &          /,' IWEAK  =',I2,'  IHVP   =',I2,'  IQCD   =',I2,
    &          '  IBOX    =',I2,/,' INTERF =',I2,'  IFINAL =',I2,
    &           '  IPHOT2 =',I2,'  IAMT4   =',I2)
1700 FORMAT (1X,'GEOMETRICAL ACCEPTANCE CUTS: ANG1,ANG2=',2(F5.1,2X))
2500 FORMAT (/,1X,'SQRT(S) [GeV]',6X,'CROSS SECTION [nb]',4X,
    &         'ASYMMETRY')
2600 FORMAT (/,1X,'SQRT(S) [GeV]',6X,'CROSS SECTION [nb]' )
1800 FORMAT (/,1X,'KINEMATICAL CUT: SPRIME=',E6.1,1X,'GEV')
1900 FORMAT (/,1X,'KINEMATICAL CUTS: ACOL=',F5.1,1X,'DEGREES',2X,
    &            'EMIN=',F5.1,1X,'GEV')
1600 FORMAT (/,1X,'NO KINEMATICAL CUT (IFAST=1)')
2000 FORMAT (1X,/,' CHOSEN FERMION CHANNEL INDEX - INDF=',I2)
3000 FORMAT (7X,F7.3,11X, F13.4)
3100 FORMAT (7X,F7.3,11X,2F13.4)
      END
\end{verbatim}
\vspace{1cm}
{\Large \bf Output of the Test Example}    \\
%For comparisons we give the result of the above quoted example.
\begin{verbatim}
 
 EXTRA Z FROM E6 GUT
 
 
 INPUT :  MZ    = 91.180 GEV    MT    = 150.00 GEV    MH  =  300.00 GEV
          AMZE  =  500.0 GEV    ZMIX  = 0.0100     TETAE6 =  0.0000
 
 QCD CORRECTION FACTORS = 1.04000  1.04500
 OUTPUT:    SW2         = .2260
 PARTIAL AND TOTAL Z-WIDTHS IN MEV
     nu,nubar  =  169.2        e+,e-  =   83.2      mu,mubar =   83.2
     tau+,tau- =   83.0        u,ubar =  296.5        d,dbar =  386.0
     c,cbar    =  296.1        s,sbar =  386.0
     t,tbar    =    0.0        b,bbar =  379.8        TOTAL  = 2501.6

 
 FLAGS IN INITIALISATION ROUTINE ZINITF:
 IWEAK  = 1  IHVP   = 3  IQCD   = 3  IBOX    = 1
 INTERF = 1  IFINAL = 1  IPHOT2 = 1  IAMT4   = 1
 
 NO KINEMATICAL CUT (IFAST=1)
 
 CHOSEN FERMION CHANNEL INDEX - INDF= 2
 
 SQRT(S) [GeV]      CROSS SECTION [nb]    ASYMMETRY
        87.000                  0.1393      -0.3713
        88.000                  0.2109      -0.2950
        89.000                  0.3671      -0.2071
        90.000                  0.7519      -0.1122
        91.000                  1.4042      -0.0192
        91.170                  1.4533      -0.0047
        92.000                  1.1713       0.0566
        93.000                  0.6813       0.1090
        94.000                  0.4338       0.1448
        95.000                  0.3076       0.1703
\end{verbatim}
%************************************************
 
\end{document}